\documentclass[aps,pre,twocolumn,superscriptaddress,floatfix,10pt]{revtex4}
\usepackage{graphicx}
\usepackage{dcolumn}
\usepackage{bm}
\usepackage{latexsym}
\usepackage{cleveref}
\usepackage{amsmath}
\begin{document}
\title{Mixed molecular motor traffic on nucleic acid tracks: models of
  transcriptional interference and regulation of gene expression{\footnote{Author names in alphabetical
      order}}}
\author{Tripti Bameta} \affiliation{CBS, University of
  Mumbai} 
\author{Debashish Chowdhury{\footnote{Corresponding author; e-mail: debch@iitk.ac.in}}}
\affiliation{Department of Physics, Indian Institute of Technology
  Kanpur, 208016}
\author{Dipanwita Ghanti} \affiliation{Department of Physics, Indian
  Institute of Technology Kanpur, 208016} \author{Soumendu Ghosh}
\affiliation{Department of Physics, Indian Institute of Technology
  Kanpur, 208016}
\begin{abstract}
RNA polymerase (RNAP) is molecular machine that polymerizes a RNA molecule, 
a linear heteropolymer, using a single stranded DNA (ssDNA) as the 
corresponding template; the sequence of monomers of the RNA is dictated by 
that of monomers on the ssDNA template. While polymerizing a RNA, the RNAP 
walks step-by-step on the ssDNA template in a specific direction. Thus, a 
RNAP can be regarded also as a molecular motor and the sites of start and 
stop of its walk on the DNA mark the two ends of the genetic message that 
it transcribes into RNA.
Interference of transcription of two overlapping genes is believed to 
regulate the levels of their expression, i.e., the overall rate of the 
corresponding RNA synthesis, through suppressive effect of 
one on the other. Here we model this process as a mixed traffic of two 
groups of RNAP motors that are characterized by two distinct pairs of  
start and stop sites. Each group polymerizes identical copies of a RNA 
while the RNAs polymerized by the two groups are different. These models, 
which may also be viewed as two interfering totally asymmetric simple 
exclusion processes, account for all modes of transcriptional interference 
in spite of their extreme simplicity. A combination of mean-field theory 
and computer simulation of these models demonstrate the physical origin 
of the switch-like regulation of the two interfering genes in both 
co-directional and contra-directional traffic of the two groups of RNAP 
motors.
\end{abstract}

\maketitle

\section{Introduction}

Synthesis of messenger RNA, a linear heteropolymer, using the corresponding 
template DNA, is called transcription; it is carried out by a molecular 
machine called RNA polymerase (RNAP) \cite{buc}.  
This machine also exploits the DNA template as a filamentous track for its
motor-like movement consuming input chemical energy \cite{gelles98}.
Polymerization of each RNA by a RNAP takes place in three stages: (a)
initiation at a specific `start' site (also called initiation site) on the 
template, (b) step-by-step elongation of the RNA, by one nucleotide in each 
forward step of the RNAP motor, and (c) termination at a specific `stop' 
site (also called termination site) on the template.  
For the sake of convenience, throughout this paper we refer to the segment 
of the template DNA between the start and the stop sites as a `gene'.

RNAPs moves from $3^{\prime}$ to the $5^{\prime}$ direction on a single strand of DNA.
Often multiple RNAPs polymerize the same gene simultaneously. In such
RNAP traffic \cite{chowdhury05,chou11}, all the RNAPs engaged 
simultaneously in the transcription process move in the same direction 
while polymerizing identical copies of a RNA, all by initiating 
transcription from the same start site and terminating at the same stop site.  
Since any segment of the template DNA covered by one RNAP is not accessible
simultaneously to any other RNAP, this steric exclusion gives rise to
nontrivial spatio-temporal organization of RNAPs in RNAP traffic.
Theoretical models of this kinetic process have been developed over
the last few years 
\cite{chou11,schad10,tripathi08,klumpp08,klumpp11,sahoo11} 
by appropriately adapting, and extending, asymmetric exclusion process 
(ASEP) \cite{schutz01,mallick15}
, a popular model in nonequilibrium
statistical mechanics that we describe briefly in the next section 

In this communication we report theoretical studies of more complex
RNAP-traffic phenomena that are believed to play important regulatory
roles in living cells \cite{pelechano13,georg11,lapidot06,kornienko13}.  
These phenomena arise from simultaneous
transcription of two overlapping genes either on the same DNA template
or two genes on the two adjacent single strands of a duplex (double-stranded)
DNA.  In the former case, traffic is entirely uni-directional although
RNAPs transcribing different genes polymerize two distinct species of
RNA molecules by starting (and stopping) at different sites on the
same template DNA strand. In contrast, in the latter case, RNAP
traffic in the two adjacent ``lanes'' move in opposite directions
transcribing the respective distinct genes.  In both these situations
the phomenon of suppressive influence of one transcriptional process
on the other is called transcriptional interference (TI)
\cite{shearwin05,mazo07}.

In general, a RNAP at the initiation, elongation or termination stage
of one transcriptional process can suppress the initiation, or
elongation (or termination) of the other transcription by another
RNAP \cite{pelechano13,georg11,lapidot06}. In other words, the stages of transcription 
of the two interfereing RNAPs define a distinct mode of interference. 
Different modes of interference have been assigned different names like
``occlusion'', ``collision'', ``sitting duck interference'',
etc. \cite{shearwin05}. 
Many pairs of interfering transcription processes are known to form 
a bistable switch: switching ON a high level of transcription of one 
of the two genes can switch OFF the other by its suppressive effect 
\cite{pelechano13,georg11,lapidot06}. 
The main aim of this communication is to demonstrate this effect 
using a unified theoretical framework that we develop here. This 
framework is capable of throwing light on many other kinetic aspects 
of TI phenomena.

In this paper we develop a unified theoretical framework that, for a 
given relative orientation of two genes, captures all possible modes 
of TI by a single set of master equations. Solving these equations, 
we investigate the effects of TI on the rates of the two transcriptions. 
Moreover, by carrying out extensive computer simulations of our model, 
we also test the validity of the mean-field approximations made in 
formulating the master equations. We explore the effects of the 
spatial extent and relative orientation of the overlap of the two genes 
as well as 
those of the kinetic parameters like initiation rates. Our results 
demonstrate interesting regulatory phenomena arising from TI. In 
particular, the transcription of one gene can be practically switched 
off by increasing that of another to a sufficiently high level.

\section{Model}

\begin{figure}[h]
  \includegraphics[angle=0,width=0.9\columnwidth]{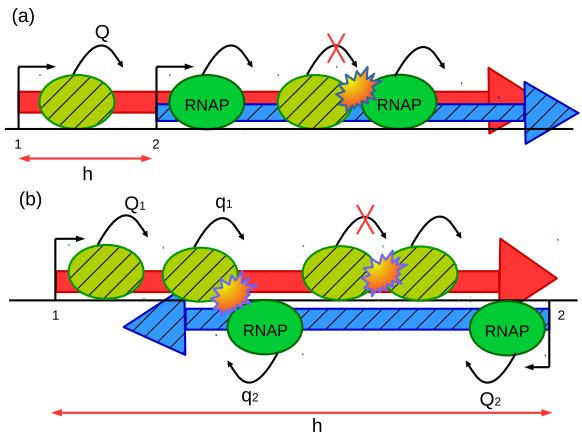}
  \caption{Schematic representation of TI model. 
(a) Codirectional mixed traffic with separate ramps: transcription 
    of two overlapping genes.
(b) Contradirectional traffic on adjoining unequal tracks: sense 
    and anti-sense transcription of two overlapping genes 
}
  \label{fig-model}
\end{figure}


We represent a single-stranded DNA (ssDNA) by a linear chain (i.e., a 
one-dimensional lattice) of equispaced sites that are labelled by the 
integer index $i$. Each site of this 
lattice denotes a nucleotide which is a monomeric subunit of the DNA track. 
The chain serves both as the template and track for the 
respective RNAP motors that are engaged in its transcription. We also  
represent each RNAP by a hard rod, i.e., an extended particle, 
of length ${\ell}$
in the units of nucleotide length, i.e., it covers ${\ell}$ successive
sites of the lattice simultaneously.  Normally, ${\ell}$ is typically
$30$ to $35$ nucleotides. We denote the position of a RNAP on its track by the 
lattice site at which the {\it leftmost} unit of the rod is located while 
the next ${\ell}-1$ sites of the lattice are merely {\it covered} by the 
RNAP. Thus, if the lattice site $j$ denotes the position of a RNAP then 
the RNAP covers not only the site $j$ but also the next ${\ell}-1$ 
sites $j+1, j+2,...,j+{\ell}-1$.   
The RNAPs interact with each
other with only hard core repulsion that is captured by imposing the
condition that no lattice site is allowed to be covered by more than
one RNAP simultaneously.

We model both codirectional TI (TTI) and contradirectional TI (CTI) 
using suitably extended TASEPs.  
For modeling 
interference of the expression of two genes we incorporate two interfering 
TASEPs each characterized by the respective distinct pair of start and 
stop sites and two distinct species of RNAP motors. 
For modeling TTI only a ssDNA template needs to be treated as the track 
for both group of RNAP motors that transcribe the two distinct genes 
simultaneously. But, for modeling simultaneous polymerization of sense 
and anti-sense transcripts, which are encoded on the two distinct strands 
of a duplex DNA, we use two antiparallel lattices on which 
the contra-directional traffic of the two groups of motors take place. 
 
We identify two sites separated by $h$ nucleotides as the start sites 
for the two genes; $h$ is an integer that can be positive, negative or 
zero (see fig.\ref{fig-model}).  A fresh initiation of transcription of a gene, however, is not 
possible as long as the first ${\ell}$ sites, starting from the 
start site of that gene, remain fully or partly covered by any other 
RNAP, irrespective of the identity of the gene that is being 
transcribed by the latter.
We denote the rates of initiation of transcription of the two genes by
$\alpha_1,\alpha_2$, respectively. 
Whenever ${\ell}$ successive sites, starting
from the start site of a gene on the DNA template is vacant, a fresh
RNAP is allowed to cover those ${\ell}$ sites thereby initiating the
corresponding transcription.
Each RNAP carries an 
unique label $1$ or $2$ depending on which of the two genes it is engaed 
in transcribing; the label is assigned to it depending on the start site 
from where it begins its walk on its track. 
Irrespective of the actual numerical value of
${\ell}$, each RNAP can move forward by only one site in each step,
provided the target site is not already covered by any other
RNAP. Single-site stepping rule is motivated by the fact that a RNAP
must transcribe the successive nucleotides one by one.
A RNAP engaged in the 
transcription of one of the two genes can detach from the lattice only 
after it reaches the stop site of the corresponding gene.  
However, so far as the rates of
termination are concerned, we assume the corresponding rates to be
$\beta_1=\beta_2=\beta$.  
$L1$ and $L2$ denote the lengths of the two genes, measured in terms of the number of 
lattice sites from the start to the stop sites of the corresponding 
gene.
The interference between the transcription of these two genes
takes place in the region of their overlap.

In the case of TTI no RNAP can pass the other immediately in front of 
it irrespective of which genes are being transcribed by the two RNAPs. 
This is motivated by the fact that in case of TTI both the genes are 
encoded on the same ssDNA strand. In contrast, in the case of CTI, 
when two RNAPs labelled by two distinct integer indices 1 and 2 (i.e., 
transcribing different genes) face each other head-on, these are allowed 
to pass, albeit with a hopping rate that is lower than that in the 
absence of the obstruction, i.e., $q_{1} < Q_{1}$ and $q_{2} < Q_{2}$. 
 This prescription is motivated by the fact 
that the genes for the sense and antisense transcripts are encoded on 
the two distinct complementary strands of the duplex DNA which serve as 
the tracks 
for the corresponding oppositely moving RNAP traffic. However, in the 
same model 
of CTI if a RNAP finds itself just behind another co-directional RNAP 
in front (i.e., both transcribe the same gene and, therefore, carry 
the same integer label) then the trailing RNAP remains stalled till 
the site in front of it is vacated by the leading RNAP. The motivation 
for this prescription is that the two RNAPs engaged in transcribing  
the same gene move on the same ssDNA strand.

\subsection{Master equations under mean-field approximation}

Let $P_{\mu}(i,t)$ denote the probability that at time $t$ there is a
RNAP at site $i$ engaged in the transcription of the gene $\mu$
($\mu=1,2$ for the genes 1 and 2, respectively). Note that the
probability that the site $i$ is occupied by a RANP, irrespective of
the gene it is transcribing, is given by $P(i) = \sum_{\mu=1}^{2}
P_{\mu}(i)$.

\subsubsection{Co-directional traffic}

Let $P(\underline{i}|j)$ be the conditional probability that, given a
RNAP at site $i$, there is another RNAP at site $j$ located downstream 
along the lattice. Obviously, $\xi(\underline{i}|j) = 1 - P(\underline{i}|j)$ 
is the conditional probability that, given a RNAP at site $i$, site $j$ is 
empty. Therefore, by definition,\\
\begin{eqnarray}
\xi(\underline{i}|i+{\ell})=\frac{1-\sum\limits_{s=1}^{{\ell}}
  P(i+s)}{1-\sum\limits_{s=1}^{{\ell}} P(i+s) + P(i+{\ell}) }\nonumber
\end{eqnarray}\\
Let $\xi(j)$ be the probability that site $j$ is not covered by any 
RNAP, irrespective of the state of occupation of any other site, is 
given by  $1-\sum_{s=0}^{{\ell}-1}P(j-s)$.
Note that, if site $i$ is given to be occupied by one
RNAP, the site $i-1$ can be covered by another RNAP if, and only if,
the site $i-{\ell}$ is also occupied.  

Under mean-field approximation, the master equations governing the 
stochastic kinetics of the two interfereing transcriptional processes 
are given by (for $h>0$)
\begin{widetext}
  \begin{eqnarray}
    \frac{dP_{1}(1,t)}{dt} &=& ~\alpha_1 \Biggl(1-\sum_{s=1}^{{\ell}}~P(s)\Biggr)
    - Q P_{1}(1,t) ~\xi(\underline{1}|1+{\ell}),~\nonumber\\
    \frac{dP_{1}(i,t)}{dt} &=&  ~Q P_{1}(i-1,t)\xi(\underline{i-1}|i-1+{\ell}) ~ - ~ Q P_1(i,t) \xi(\underline{i}|i+{\ell}) ~~~{\rm ~for~}, ~(1<i<L1)~, \nonumber \\
    \frac{dP_{1}(L1,t)}{dt} &=&  ~Q P_{1}(L1-1,t)\xi(\underline{L1-1}|L1-1+{\ell}) - \beta P_1(L1,t),\nonumber\\
    \frac{dP_{2}(1+h,t)}{dt} &=& ~\alpha_2\underbrace{\Biggl(1-\sum_{s=0}^{{\ell-1}}~P_{1}(h-s)\Biggr)}_{=1 \rm ~for~(h = 0)} \Biggl(1-\sum_{s=1}^{{\ell}}~P(h+s)\Biggr) - Q P_{2}(1+h,t) ~\xi(\underline{1+h}|1+h+{\ell}), \nonumber \\
    \frac{dP_{2}(i,t)}{dt} &=&  ~Q P_{2}(i-1,t)\xi(\underline{i-1}|i-1+{\ell}) ~ - ~ Q P_2(i,t) \xi(\underline{i}|i+{\ell})~~{\rm~ ~for~ },~ (1+h<i<h+L2), \nonumber \\
    \frac{dP_{2}(h+L2,t)}{dt} &=&  ~Q P_{2}(h+L2-1,t)\xi(\underline{h+L2-1}|h+L2-1+{\ell}) - \beta P_2(h+L2,t).\nonumber\\
    \label{eq-obc6}
\end{eqnarray}
\end{widetext}
Equations for $h<0$ can be obtained from (\ref{eq-obc6}) by interchanging index 1 and 2.
\subsubsection{Contra-directional traffic}

Effect of CTI on gene expression can be studied by writing down and
solving master equation for $P_{\mu}(i,t)$, which denotes the probability
of finding an RNAP on gene $\mu (\mu \equiv {1,2})$ at time t, at lattice
site $i$. Obviously, $\xi_{1}(\underline{i}|i+\ell)$ is the conditional probability that, given a RNAP  on gene 1 at site $i$, site $i+\ell$ is empty. Therefore, by definition,\\
\begin{eqnarray}
\xi_{1}(\underline{i}|i+\ell)&=&\frac{1 - \sum\limits_{s=1}^{\ell} P_1(i+s)}{1 - \sum\limits_{s=1}^{\ell} P_1(i+s)+P_1(i+\ell)}\nonumber
\end{eqnarray}\\
Similarly, $\xi_{2}(i-\ell|\underline{i})$ is the conditional probability that, given a RNAP on gene 2 at site $i$ , site $i-\ell$ is empty. Therefore, by definition,\\
\begin{eqnarray}
\xi_{2}(i-\ell|\underline{i})&=&\frac{1-\sum\limits_{s=1}^{\ell}
  P_2(i-s)}{1-\sum\limits_{s=1}^{\ell} P_2(i-s)+P_2(i-\ell)} \nonumber
\end{eqnarray}\\
Let $\xi_{1}(j)$ be the probability that site $j$ on gene 1 is not covered 
by any RNAP, irrespective of the state of occupation of any other site. 
obviously,  $\xi_{1}(j) =  1-\sum_{s=0}^{{\ell}-1}P_{1}(j-s)$.
Similarly, $\xi_{2}(j)$, the probability that site $j$ on gene 2 is not 
covered by any RNAP, irrespective of the state of occupation of any other 
site, is given by  $1-\sum_{s=0}^{{\ell}-1}P_{2}(j-s)$.

Under mean-field approximation, the master equations are written as
\begin{widetext}
\begin{eqnarray}
  \frac{d P_1 \left(1,t\right)}{dt}&=&\alpha_1 \left[1-\sum\limits_{s=1}^{\ell}P_1\left(s,t \right) \right] \underbrace{\left[1-\sum\limits_{s=1}^{\ell}P_2\left(s,t \right) \right]}_{=1 \rm~ for ~(h-L2>\ell)} 
 - P_1(1,t)\xi_{1}(\underline{1}|1+{\ell})\left[Q_1\xi_2(1+\ell)+q_1\{1-\xi_2(1+\ell)\}\right],\nonumber \\
  \frac{d P_1\left(i,t\right)}{dt}&=&P_1(i-1,t)\xi_{1}(\underline{i-1}|i-1+{\ell})\left[Q_1\xi_2(i-1+\ell)+q_1\{1-\xi_2(i-1+\ell)\}\right] \nonumber \\
  &-& P_1(i,t)\xi_{1}(\underline{i}|i+{\ell})\left[Q_1\xi_2(i+\ell)+q_1\{1-\xi_2(i+\ell)\}\right]~~~{\rm ~for~}, ~(1<i<L1)~,\nonumber \\
  \frac{d P_1\left(L1,t\right)}{dt}&=&P_1(L1-1,t)\xi_{1}(\underline{L1-1}|L1-1+{\ell})\left[Q_1\xi_2(L1-1+\ell)+q_1\{1-\xi_2(L1-1+\ell)\}\right] - \beta P_1\left(L1,t\right),\nonumber \\
  \frac{d P_2 \left(1+h,t\right)}{dt}&=&\alpha_2 \underbrace{\left[1-\sum\limits_{s=0}^{\ell-1}P_1\left(1+h-s,t \right) \right]}_{=1 \rm~ for ~(h-L1>\ell)}  \left[1-\sum\limits_{s=0}^{\ell-1}P_2\left(1+h-s,t \right) \right] \nonumber \\ 
  &-& P_2(1+h,t)\xi_{2}(1+h-{\ell}|\underline{1+h})\left[Q_2\xi_1(h)+q_2\{1-\xi_1(h)\}\right],\nonumber \\
   \frac{d P_2\left(i,t\right)}{dt}&=&P_2(i+1,t)\xi_{2}(i+1-{\ell}|\underline{i+1})\left[Q_2\xi_1(i)+q_2\{1-\xi_1(i)\}\right] \nonumber \\ 
  &-& P_2(i,t)\xi_{2}(i-{\ell}|\underline{i})\left[Q_2\xi_1(i-1)+q_2\{1-\xi_1(i-1)\}\right]~~~{\rm ~for~}, ~(2+h-L2<i<1+h)~, \nonumber \\
  \frac{dP_2\left(2+h-L2,t\right)}{dt}&=&P_2(3+h-L2,t)\xi_{2}(3+h-L2-{\ell}|\underline{3+h-L2})\left[Q_2\xi_1(2+h-L2)+q_2\{1-\xi_1(2+h-L2)\}\right]\nonumber \\ 
&-& \beta
  P_2\left(2+h-L2,t\right). \nonumber \\ 
\label{eqs:6}
\end{eqnarray}
\end{widetext}

\section{Results}

Solving the master equations (\ref{eq-obc6}) and (\ref{eqs:6}) numerically 
under steady 
state conditions we obtained the corresponding rates of the transcriptions 
of the two genes. Moreover, in order to test the range of validity of the 
MFA made in writing the master equations, we also carried out extensive 
direct computer simulations of our model using the same set of parameter 
values that we used for solving the master equations. During the simulations, 
we monitored the flux of the RNAPs. The system needed, typically, about 
two million time steps to attain the steady state after which we collected 
the steady-state data over the next five million time steps. The steady-state 
properties presented in this paper are averages of the data collected only 
in the steady state of the system.  All the numerical results plotted in 
this paper have been obtained for ${\ell}=10$, $L_1=1000$ and $L_2=1100$;  
by comparing with the results for a few other lengths of RNAPs and genes, 
we ensured that our conclusions do not suffer from any artefacts of the 
choice of these parameters.

\subsection{Results for co-directional traffic}

\begin{figure}[t]
  \begin{center}
    \includegraphics[width=0.95\columnwidth]{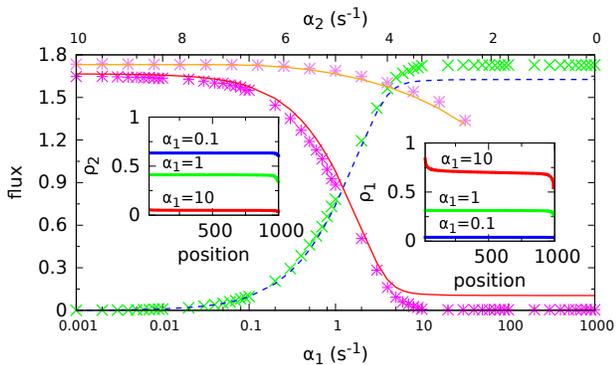}\\[0.25cm]
  \end{center}
  \caption{Codirectional: The switch like behavior of fluxes of RNAPs, plotted as a function of ${\alpha}_{1}$, for $Q=30 ~s^{-1}$, ${\alpha}_{2}=5 ~s^{-1}$, $\beta_1 = \beta_2 = 1000 S ^ {-1}$ and $h=20 ~bp$.The dashed line and solid line corresponds to our mean-field theoretic predictions for flux 1  and flux 2 respectively
whereas the discrete data points (cross and star corresponds for flux 1 and flux 2 respectively) have been obtained from computer simulations.The insets show the average density profiles for three different values of ${\alpha}_{1}$ .When the gene does not have a transcriptional interference, its expression follows different kinetics.Corresponding flux is plotted as a function of ${\alpha}_{2}$, for $Q=30 ~s^{-1}$ ($\alpha_{1}=0$).The solid orange line corresponds to our mean-field theoretic predictions and violet data points have been obtained from computer simulations.
}
  \label{fig-switchCO}
\end{figure}

\begin{figure}[t]
  \begin{center}
    \includegraphics[width=0.95\columnwidth]{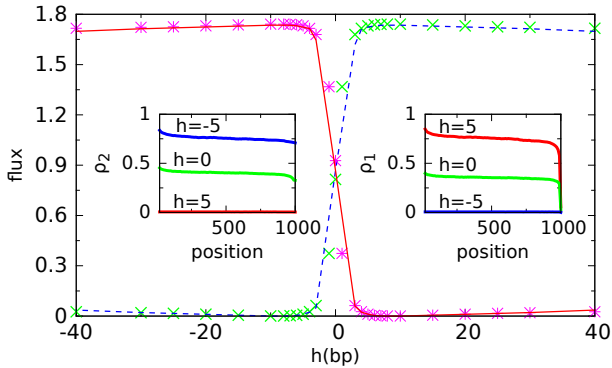}\\[0.25cm]
  \end{center}
  \caption{Codirectional: The roles of the two genes as ``suppressor'' 
and ``suppressed' are interchanged as $h$ is varied from -N to +N (N=40 
in this figure just for illustration).
}
  \label{fig-interch}
\end{figure}

In fig.\ref{fig-switchCO} we plot the fluxes of RNAPs in the two traffic 
which, as explained above, are the overall rates of transcription of the 
two genes. First of all, note that for any given value of $\alpha_2$, 
gene 2 would have got transcribed normally at a fairly high rate if the 
gene 1 were not interfering with its transcription. As long as $\alpha_1$ 
is not too high, the rate of transcription of gene 2 is weakly affected 
by infrequent co-directional ``collisions'' and proceeds at a fairly high 
rate. But, as $\alpha_1$ increases the time gap detected at any arbitrary 
site between the departure of a RNAP and the arrival of the next RNAP 
becomes shorter. Therefore, the site for the initiation of transcription 
of gene 2, which is located on the path of the RNAP traffic on gene 1, 
remains ``occluded'' for most of the time if $\alpha_1$ is sufficiently 
high. Consequently, a high rate of expression of gene 1 strongly suppresses 
the expression of gene 2, irrespective of the actual numerical value of 
$\alpha_2$. Thus, the rates of transcription of the two genes are strongly 
anti-correlated. 

The role of ``suppressor'' and suppressed'' genes are interchanged as 
the separation $h$ between the transcription initiation sites is varied 
from a positive integer to a negative integer (see fig.\ref{fig-interch})
The sharp changes take place only over a narrow interval of the order of 
${\ell}$ around $h=0$.

\subsection{Results for contradirectional traffic}

\begin{figure}[h]
  \includegraphics[width=0.8\columnwidth]{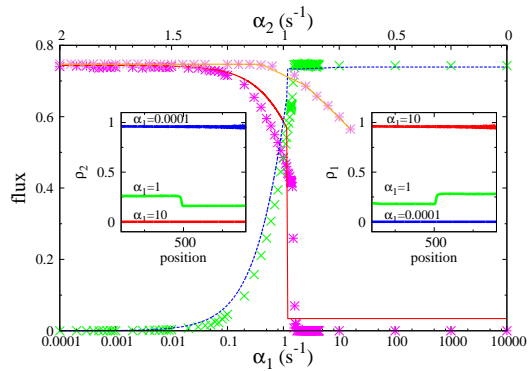}
  \centering
  \caption{Contradirectional: Same as in fig.\ref{fig-switchCO}, except that the traffic is contradirectional and $h=500$, $Q_{1}=Q_{2} = 30 ~s^{-1}$, $q_{1} = q_{2} = 10 ~s^{-1}$, ${\alpha}_{2}=1.0 ~s^{-1}$, $\beta_{1}=\beta_{2}=1.0 s^{-1}$, $L1=L2=1000$. 
}
\label{fig-switchOPP}
\end{figure}

Although the results on flux, plotted in fig.\ref{fig-switchOPP}, are 
qualitatively similar to the corresponding results for co-directional 
traffic plotted in fig.\ref{fig-switchCO}, there are some additional 
features. The kinks observed in the density profiles shown in the 
insets of fig.\ref{fig-switchOPP} are consequences of the extended 
``defect'' created by the slower moving RNAPs against the faster moving 
ones \cite{sinha15}.

\section{Summary and conclusion}

Co-directional and contra-directional two-species TASEP, both on a single 
track and two parallel tracks, have been studied earlier for purely 
theoretical consideration as well as for capturing real physical processes 
\cite{schutz03,kunwar06,john04,lin11}. 
But, in all those models a single pair of start and stop sites serve as 
the entry and exit points for both species of particles. Motivated by cytoskeletal motor traffic, TASEP-based models with two 
distinct species of oppositely moving self-propelled particles have  
been developed earlier \cite{chai09,ebbinghaus09,ebbinghaus10,muhuri10,neri13}. 
Since those motors can attach to and detach from any lattice site, 
except for quantitative difference in the values of attachment/detachment, 
the motor kinetics at the two ends of the track were qualitatively no 
different from those at any other site. In contrast, for the transcription 
of a specific gene, RNAP motors have to start and stop their walk at 
pre-designated sites. Moreover, premature detachment of a RNAP, which would 
produce a truncated RNA strand, is not allowed in our model because such 
errors occur very rarely. 

In most of the earlier theoretical models on RNAP traffic 
\cite{tripathi08,klumpp08,klumpp11,sahoo11,ohta11} 
all the RNAP were engaged in transcribing a single gene; therefore the 
traffic was uni-directional and every RNAP polymerized identical copies 
of the RNA while a single pair of start-stop sites marked the points 
of initiation and termination of transcription. In contrast, in the 
models reported in this paper two distinct pairs of start-stop sites 
mark the points of initiation and termination of the respective genes. 
Moreover, the RNA species that gets polymerized by a RNAP depends on 
the sites from which it initiates transcription. Thus copies of two 
distinct species of RNA get synthesized simultaneously by a mixed 
population of two groups of RNAP motors the relative direction of whose 
movements is dictated by the relative orientation of the two genes.  
 
In our model of TI in co-directional mixed RNAP traffic we assumed that 
a RNAP passively waits at its current position if the target nucleotide 
in front is already covered by another RNAP. The temporarily stalled 
RNAP can resume its forward movement, and the concomitant transcriptional 
activity, only after its target site is vacated by the RNAP immediately 
in front of it. We have also ignored the possibility of backtracking of 
the individual RNAP motors \cite{nudler12,landick06,sahoo11}. 
In future extensions of our model \cite{ghosh15}, we intend to explore 
the effects of backtracking, active re-starting of stalled RNAP by a 
trailing RNAP \cite{epshtein03a,dong12} as well as premature detachment 
of RNAPs upon suffering collision.  

In addition to the differences in the models developed here and all 
other TASEP-type models described above, the main questions addressed 
in those models are also fundamentally different from those addressed  
in this paper. 
In all the earlier theoretical works the
effects of the different modes of interference have been studied
separately \cite{sneppen05,palmer09}. 
The simple TASEP-based unified model that we have developed here not 
only captures all possible modes of transcriptional interference, but 
also accounts for the self-regulation of the pair of genes through 
the adaptation of the levels of their interfering transcriptions. 

\section*{Acknowledgements} 
This work is supported by a J.C. Bose National Fellowship (DC) and DST 
Inspire Faculty Fellowship (TB).


\end{document}